\documentclass[final,3p,times,twocolumn]{elsarticle}
\biboptions{comma,sort&compress}
\usepackage{ecrc}
\usepackage{here}
\usepackage{graphicx}
\usepackage{epsfig}
\usepackage{epstopdf}
\usepackage{amsmath}
\usepackage{lipsum}

\def\nin{\noindent}
\def\beq{\begin{equation}}
\def\eeq{\end{equation}}
\def\bea{\begin{eqnarray}}
\def\eea{\end{eqnarray}}
\def\nnb{\nonumber}
\def\la{\langle}
\def\ra{\rangle}

\usepackage{graphicx}
\usepackage{here}
\def\beq{\begin{equation}}
\def\eeq{\end{equation}}
\def\bea{\begin{eqnarray}}
\def\eea{\end{eqnarray}}
\def\bq{\begin{quote}}
\def\eq{\end{quote}}
\def\ve{\vert}
\def\nnb{\nonumber}

\def\nnb{\nonumber}
\def\la{\langle}
\def\ra{\rangle}
\def\nin{\noindent}
\def\ba{\begin{array}}
\def\ea{\end{array}}

\def\als{\alpha_s}

\def\gg2{ \la\alpha_s G^2 \ra}
\def\gg3{g^3f_{abc}\la G^aG^bG^c \ra}
\def\ggg4{\la\als^2G^4\ra}

\def\beq{\begin{equation}}
\def\enq{\end{equation}}
\def\beqa{\begin{eqnarray}}
\def\enqa{\end{eqnarray}}
\def\nnb{\nonumber}

\def\MeV{\nobreak\,\mbox{MeV}}
\def\GeV{\nobreak\,\mbox{GeV}}
\def\keV{\nobreak\,\mbox{keV}}

\newcommand{\rag}{\rangle}
\newcommand{\lag}{\langle}

\def\gg{\lag g^{2}_{s} G^2 \rag}
\def\ggg{\lag g^{3}_{s}G^3\rag}

\volume{00}
\firstpage{1}
\journalname{Nuclear and Particle Physics Proceedings }
\runauth{D. Rabetiarivony}
\jnltitlelogo{Nuclear and Particle Physics Proceedings }

\begin{document}
\begin{frontmatter}

\title{Light scalar quarkonia from Laplace sum rule at NLO\tnoteref{text1}}

\author[label1]{R. M. Albuquerque}
\ead{raphael.albuquerque@uerj.br}
\address[label1]{Faculty of Technology,Rio de Janeiro State University (FAT,UERJ), Brazil}
\author[label2,label3]{S. Narison}
\ead{snarison@yahoo.fr}
\address[label2]{Laboratoire
Univers et Particules de Montpellier (LUPM), CNRS-IN2P3, Case 070, Place Eug\`ene
Bataillon, 34095 - Montpellier, France}
\address[label3]{Institute of High-Energy Physics of Madagascar (iHEPMAD), University of Antananarivo, Antananarivo 101, Madagascar}


\author[label3]{D. Rabetiarivony\fnref{fn2}}
\fntext[fn2]{Speaker}
\ead{rd.bidds@gmail.com} 


\tnotetext[text1]{Talk given at QCD24 International Conference (8--12 July 2024, Montpellier--FR)}

\pagestyle{myheadings}
\markright{ }

\begin{abstract}
\noindent
We review our results on light scalar quarkonia ($\bar{q}q$ and four-quark states) from (inverse) QCD Laplace sum rules (LSR) and their ratios ${\cal R}$ within stability criteria and including higher order perturbative (PT) corrections up to the (estimated) ${\cal O}(\alpha_{s}^{5})$. As the Operator Product Expansion (OPE) usually converges for $D\leqslant 6-8$, we evaluated the QCD spectral functions at Lowest Order (LO) of PT QCD and up to the $D=6$ dimension vaccum condensates. We request that the optimal results obey the constraint: Pole (Resonance) contribution to the spectral integral is larger than the QCD continuum one which excludes an on-shell mass around $(500-600)~\MeV$ obtained for values of the QCD continuum threshold $t_c\leqslant(1\sim 1.5)~\GeV^2$. Our results for the different assignments of the scalar mesons are compiled in Tables\,\ref{tab:resqqmol} to \ref{tab:resva}.
\end{abstract} 
\scriptsize
\begin{keyword}
QCD Spectral Sum Rules \sep Perturbative and Non-perturbative QCD \sep Exotic hadrons \sep Scalar mesons \sep Masses and Decay constants.
\end{keyword}
\end{frontmatter}
\section{Introduction}
This talk summarizes the results of our original works in Ref.\,\cite{ANRls}.

\nin It is known that the underlying structure of the light scalar mesons is not well established both from theory and from experiments. Several different pictures have been proposed for these states. Among different theoretical interpretations of the light scalar mesons we have: ordinary mesons $\bar{q}q$\,\cite{SNB4,SNA,MNP,NPRT,SND,RRY,SNB5,BSN}, four-quark states and $\pi^+\pi^-,\,K^+K^-,\,K\pi,\cdots$ molecules\,\cite{JAFFE,ADS,IW,SNB2,SNB5,LP,BNNB,CHZ,CS}. In this paper, we present improved analysis of the determinations of the masses and couplings of ordinary $\bar{q}q$ and four-quark states, and new results for $\pi^+\pi^-,\,K^+K^-,\cdots$ molecules using the inverse Laplace sum rule\,\cite{BELL,BELLa,BNR,BERT,NEUF,SNR,SNB1,SNB2} version of QCD spectral sum rules (QSSR)\footnote{For reviews, see \cite{SVZa,Za,SNB1,SNB2,SNB3,CK,YND,PAS,RRY,IOFF,DOSCH,SNqcd22} and references therein}. 
\section{The Laplace sum rule}
\nin We shall work with the inverse transform Laplace sum rule and their ratios for extracting the decay constant and the mass:
\bea
\hspace*{-0.2cm} {\cal L}^c_n(\tau,\mu)&\equiv & \int_{t_>}^{t_c} dt~t^n~e^{-t\tau}\frac{1}{\pi} \mbox{Im}~\psi_{S}(t,\mu)~;\nnb \\
 {\cal R}^c_{10}(\tau)&\equiv &\frac{{\cal L}^c_{1}} {{\cal L}^c_0},
\label{eq:lsr}
\eea
The spectral function $\mbox{Im}~\psi_S(t,\mu)$ can be evaluated from the two-point correlator:
\bea
\hspace*{-0.6cm}
\psi_S(q^2)\hspace*{-0.2cm}&=& \hspace*{-0.2cm} i \int \hspace*{-0.1cm} d^4 x\, e^{-i q x}\lag 0 \ve \mathcal{T} \mathcal{O}_S(x)(\mathcal{O}_S(0))^{\dag} \ve 0 \rag,
\eea
where $\mathcal{O}_S(x)$ are the quark operators describing the ordinary $\bar{q}q$ mesons, $(\bar{q}q')(\bar{q}'q)$ molecules or $(\bar{q}\bar{q}')(q'q)$ four-quark states; $t_>$ is the quark threshold and $\tau$ is the LSR variable; $\mu$ is an arbitrary perturbative (PT) subtraction constant which is equal to $1/\sqrt{\tau}$ as we shall work with a renormalization group resumed PT series; $t_c$ is the threshold of the "QCD continuum" which parametrizes, from the discontinuity of the Feynman diagrams, the spectral function. Contrary to some intuitive claims in the literature, $\sqrt{t_c}$ does not necessarily coincide with the mass of the first radial excitation but can be higher as the QCD continuum smears all higher state contributions.
\section{Optimization criteria}
Like in our previous works\,\cite{LNSR,ANRls,ANRTm,ANR21,ANR22p,ANR22pa,ANRR1,ANRR1a,ANRR2,ANR1,NR1,SNX1,SNX2,SU3}, we shall use stability criteria on the external variables $\tau$ and $t_c$ for extracting the optimal results from the sum rules. These stability regions manifest either as plateau, minimum or inflexion points. We shall also request that the Pole (P) contribution (resonance) to the spectral integral is larger than the QCD continuum (C) one:
\beq
\hspace*{-0.2cm}
R_{P/C}\equiv \frac{\mbox{Lowest Pole}}{\mbox{QCD continuum}}= \frac{\int^{tc}_{t_>}dt e^{-t\tau}\mbox{Im}\psi_S (t)}{\int^{\infty}_{t_c}dt e^{-t\tau}\mbox{Im}\psi_S (t)}\geqslant 1.
\eeq
Set of $(\tau,t_c)$ stability not satisfying this requirement will be rejected. 
\section{The ordinary $\bar{q}q$ states}
We shall be concerned with the quark operators:
\bea
J_2&=&\frac{m_q}{\sqrt{2}}(\bar{u}u+\bar{d}d),\nnb\\
J_{\bar{u}s}&\equiv& \partial_{\mu}V^{\mu}_{\bar{u}s}=(m_u-m_s)\bar{u}s,\nnb \\
J_3&=&m_s\bar{s}s,
\eea
where $m_q=(m_u+m_d)/2$.

\nin As the behaviours of the different curves are similar, we shall illustrate the analysis in the case of the scalar-isoscalar $(\bar{u}u+\bar{d}d)$. The final results are compiled in Table \ref{tab:resqqmol}.
\subsection{The light scalar-isoscalar $\frac{1}{\sqrt{2}}(\bar{u}u+\bar{d}d)$ state}
We improve the estimation in \cite{SND,NPRT,RRY,SNB2,SNB5,BSN} by adding higher order PT corrections\,\cite{ANRls}. The mass corrections and  non-perturbative contributions up to $D=6$ are listed in Ref.\cite{SNB1,NPRT,JM,SNB6}.
\subsection*{$\lozenge$ Estimate of the mass and coupling}
\begin{figure}[hbt]
\begin{center}
\includegraphics[width=6.0cm]{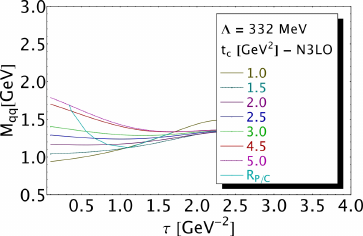}
\includegraphics[width=6.0cm]{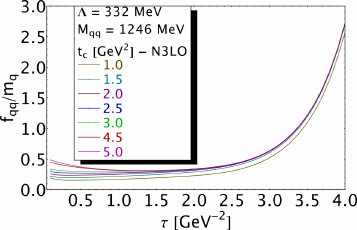}
\caption{\footnotesize The mass and coupling of $\frac{1}{\sqrt{2}}(\bar{u}u+\bar{d}d)$ as a function of $\tau$ at N3LO for \# values of $t_c$.} 
\label{fig:uu}
\end{center}
\end{figure}

\nin We study the behaviour of the coupling from the moment ${\cal L}^{c}_{0}$ and mass from the ratio of moments ${\cal R}^{c}_{10}$ in term of the LSR variable $\tau$ for different values of $t_c$ at N3LO as shown in Fig.\,\ref{fig:uu}.
\subsection*{$\lozenge$ Truncation of the PT series}
\begin{figure}[hbt]
\begin{center}
\includegraphics[width=6.0cm]{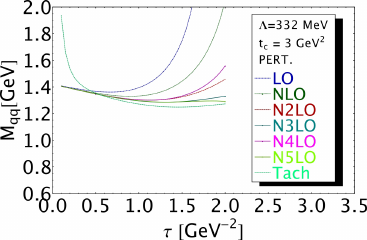}
\includegraphics[width=6.0cm]{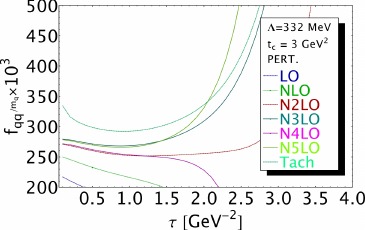}
\caption{\footnotesize The mass and coupling of $\frac{1}{\sqrt{2}}(\bar{u}u+\bar{d}d)$ as a function of $\tau$ for different truncation of the PT series for fixed value of $t_c=3\,\GeV^2$.} 
\label{fig:qqPT}
\end{center}
\end{figure}

\begin{figure}[hbt]
\begin{center}
\includegraphics[width=6.0cm]{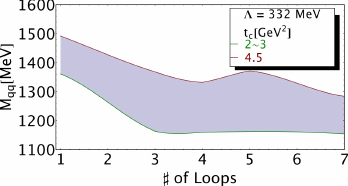}
\includegraphics[width=6.0cm]{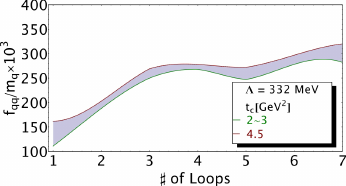}
\caption{\footnotesize The mass and coupling of $\frac{1}{\sqrt{2}}(\bar{u}u+\bar{d}d)$ for different truncation of the PT series and for two extremal values of $t_c$ where the $\tau$-stability is reached. The last point \#7 in the loop axis corresponds to the effect of tachyonic gluon mass.} 
\label{fig:qqloop}
\end{center}
\end{figure}

\nin In Fig.\,\ref{fig:qqPT}, we show the $\tau$ behaviour of the mass and coupling for a given value of $t_c=3\,\GeV^2$ for different truncation of the PT series. We have also in Fig.\,\ref{fig:qqloop} the behaviour of the mass and coupling for different truncation of the optimal results for given two extremal values of $t_c$. One can notice that the value of the coupling versus $\tau$ is almost stable. However, the higher order PT corrections shift the position of the minima to larger values of $\tau$ for the mass. One can also notice that a stability in $\tau$ appears from N2LO. We extract the optimal results at N3LO and consider as sources of the error due to the truncation of the PT series the sum N4LO$\oplus$N5LO (estimated by assuming that the coefficients of the series have a geometrical growth behaviour\,\cite{SNZ}). We obtain at the optimal value of $\tau \simeq 1\,\GeV^{-2}$:
\bea
\Delta M_{\bar{q}q}\vert_{\mbox{Geom}}&=&\pm 1\MeV,\nnb \\
\Delta f_{\bar{q}q}/m_q\vert_{\mbox{Geom}}&=&\pm 15\times 10^{-3}.
\eea
An alternative way to estimate the error is to parametrize the contribution of the non-calculated term by the tachyonic gluon mass\,\cite{ZAK,SNBs,ANDREEV,AZAK,JSNR,CSNZ}. However, as the tachyonic gluon mass tends to overestimate the error in some cases, we shall keep, here and in the following, the most optimistic errors from the geometric growth of the $\alpha_s$-coefficient.
\subsection*{$\lozenge$ Factorization of the four-quark condensate}
\begin{figure}[hbt]
\begin{center}
\includegraphics[width=6.0cm]{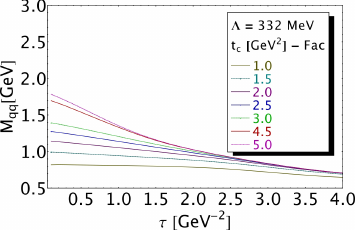}
\includegraphics[width=6.0cm]{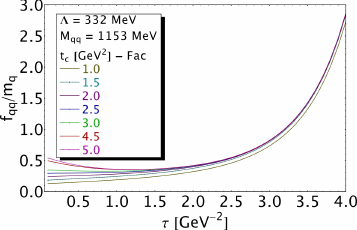}
\caption{\footnotesize $\tau$ behaviour of the mass and coupling of $\frac{1}{\sqrt{2}}(\bar{u}u+\bar{d}d)$ for different values of $t_c$ using a factorization of the four-quark condensate.} 
\label{fig:qqfac}
\end{center}
\end{figure}

\nin We shall test the effect of the vacuum saturation or factorization assumption on the mass and coupling predictions. From Fig.\,\ref{fig:qqfac}, one can notice that the $\tau$-stability of the mass is less good (an inflexion point) than in the case of a violation of factorization (Fig.\,\ref{fig:uu}) and the range of $t_c$-values is more restricted for $t_c=(3-4.5)\,\GeV^2$. However, for the coupling, the minimum appears around $\tau\sim 0.9$ and $1.2\,\GeV^{-2}$. The errors due to $t_c$ and $\tau$ are respectively $33\,\MeV$ and $77\,\MeV$ for the mass and $12\times 10^{-3}$ and $3\times 10^{-3}$ for the coupling.
\subsection*{$\lozenge$ Final results}
\nin The final and conservative results are extracted at N3LO assuming a violation of the factorization assumption. The stability region is delimited in the range $(t_c,\tau)=(2.0, 0.9) \sim (4.5, 1.5)\,(\GeV^2,\GeV^{-2})$ where, we obtain:
\beq
\hspace*{-0.1cm}
M_{\bar{q}q} = 1246(96)\MeV,~f_{\bar{q}q}/m_q (\tau)= 274(43)\times 10^{-3}.
\label{eq:fresqq}
\eeq
The different sources of the errors are quoted in Table\,\ref{tab:resqqmol}
\subsection*{$\lozenge$ Finite width correction}
\nin The previous analysis are done using a Narrow Width Approximation (NWA). In order to study the effect of the width, we use the ratio:
\beq
\left( M^{BW}_{\bar{q}q}\right)^2=\frac{\int_{0}^{t_c}~dt~t^2~e^{-t\tau}BW(t)}{\int_{0}^{t_c}~dt~t~e^{-t\tau}BW(t)},
\eeq
where, 
\beq
BW(t)=\frac{M_{\sigma}\Gamma_{\sigma}}{(t-M^{2}_{\sigma})^2+M^{2}_{\sigma}\Gamma^{2}_{\sigma}}
\eeq
and $M_{\sigma}$ is the mass from the NWA (Eq.\,\ref{eq:fresqq}) obtained for $(t_c,\tau)=(2.56, 1.05)\,(\GeV^2,\GeV^{-2})$.

\begin{figure}[hbt]
\begin{center}
\includegraphics[width=6.0cm]{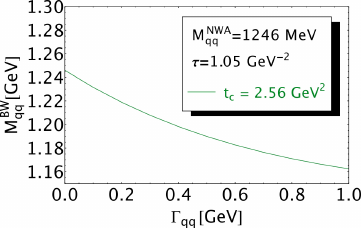}
\caption{\footnotesize The Finite width effect on the mass for $t_c=2.56\,\GeV^2$ and $\tau=1.05\,\GeV^{-2}$ corresponding to the central value of $M_{\bar{q}q}=1246\,\MeV$ in a NWA.} 
\label{fig:qqfwa}
\end{center}
\end{figure}

\nin As one can notice from Fig.\,\ref{fig:qqfwa}, varying the value of $\Gamma_{\sigma}$ from $\Gamma_{\pi\pi}=120\,\MeV$ (from vertex sum rule\,\cite{SNB4,SNB2,BSN}) to $520\,\MeV$ (for the complex pole mass\,\cite{ANRls}) and $700\,\MeV$ (for the On-shell mass\,\cite{ANRls}) the width decreases slightly the mass (in units of $\MeV$):
\beq
\hspace*{-0.15cm}
\Delta M^{BW}_{\bar{q}q}=-22\vert_{\mbox{Vertex SR}},~-60\vert_{\mbox{Pole}},~-70\vert_{\mbox{On-shell}}.
\eeq
\section{The molecule states: $\pi^+\pi^-\,,K^+K^-\,,K^+\pi^-\,,\eta\pi^0$}
We shall work with the following interpolating currents:
\bea
{\cal O}_{\pi^+\pi^-}(x)&=&(m_u+m_d)^2(\bar{d}\,i\gamma_5\, u) \otimes (\bar{u}\, i\gamma_5\,d)(x),\nnb\\
{\cal O}_{K^+K^-}(x)&=&(\bar{s}\,i\gamma_5\, u)\otimes(\bar{u}\,i\gamma_5\, s)(x),\nnb\\
{\cal O}_{K^+\pi^-}(x)&=&(\bar{s}\,i\gamma_5\, u)\otimes(\bar{u}\,i\gamma_5\, d)(x),\nnb\\
{\cal O}_{\eta\pi^0}(x)&=&\hspace*{-0.3cm}\frac{1}{\sqrt{6}}\left[(\bar{u}\,i\gamma_5 u)+(\bar{d}\,i\gamma_5 d)-2(\bar{s}\,i\gamma_5 s)\right]\nnb\\
&&\otimes \frac{1}{\sqrt{2}}\left[(\bar{u}\,i\gamma_5 u)+(\bar{d}\,i\gamma_5 d)\right](x).
\eea
\nin We do not consider the scalar $\bar{q}q'$ current which cannot contribute to leading order to the decay $\sigma\rightarrow \pi^+\pi^-$. We shall also omit the global factor $(m_u+m_d)$ of the pion current such that the corresponding two-point function has an anomalous dimension. This procedure will not affect the estimation of the mass but the coupling.

\nin The analysis is similar to the case of the ordinary $\frac{1}{\sqrt{2}}(\bar{u}u+\bar{d}d)$ and will not be repeated in this section. The final and conservative results with the different sources of the errors are given in Table\,\ref{tab:resqqmol}.
\section{Comments on $\bar{q}q$ and molecule states}
\nin -- The masses of the molecule states are about $(230-250)\MeV$ lower than the corresponding $\bar{q}q$ states. However, with the width corrections which act with opposite signs in the two cases, the two estimations tend to meet around $1.1\,\GeV$.\\
-- The predicted masses of the $\bar{q}q$ meson and molecule states coincide with the one of the lightest scalar gluonium\,\cite{SNA}.\\
-- At this level, one cannot yet distinguish the $\bar{q}q$, $\pi^+\pi^-$ molecule and gluonium nature of the $\sigma$.
\section{The four-quark states: $\overline{ud}ud\,,\overline{ud}us\,,\overline{us}ds$}
In the literature many diquark-anti-diquark configurations can describe the four-quark scalar states\,\cite{JAFFE,BNNB,CHZ,CS}. In general, the physical state should be their combination with arbitrary mixing parameters which is almost impossible to control. Among these different configurations, we choose to work with the combination "Scalar $\oplus$ Pseudoscalar" and "Vector $\oplus$ Axial-vector" currents:
\bea
\hspace*{-0.7cm}
{\cal O}^{S/P}&=&\epsilon_{abc}\epsilon_{dec}\left[\left(\bar{u}_a\gamma_5\,C\,\bar{q}_{b}^{T}\right)\otimes \left(q'^{T}_{d}\,C\,\gamma_5\,q_e\right)\right.\nnb\\
&&+r\left.\left(\bar{u}_a\,C\,\bar{q}^{T}_b\right)\otimes\left(q'^{T}_{d}\,C\,q_e\right)\right],\nnb
\eea
\bea
\hspace*{-0.7cm}
{\cal O}^{V/A}&=&\hspace*{-0.25cm}\frac{1}{\sqrt{2}}\left[\left(\bar{u}_a\gamma_{\mu}\gamma_5 C \bar{q}^{T}_{b}\right)\left(q'^{T}_{a}C\gamma^{\mu}\gamma_5 q_b-q'^{T}_{b} C \gamma^{\mu}\gamma_5 q_a \right) \right.\nnb\\
&+&\hspace*{-0.3cm}r\left.\left(\bar{u}_a\gamma_{\mu} C \bar{q}^{T}_{b}\right)\left(q'^{T}_{a} C \gamma^{\mu} q_b+q'^{T}_{b} C \gamma^{\mu} q_a \right)\right],
\eea
where, $r$ is an arbitrary mixing parameter which we shall leave as free inside the range $0$ to $1$; $q\equiv d,s$ and $q'\equiv u,d$.

\nin The analysis is similar to the previous states. The results are given in the Tables\,\ref{tab:ressp} and \ref{tab:resva}.
\section{First radial excitations}
We extend the study to the case of the first radial excitations. Observing that the $SU(3)$ breaking effects cannot be seen within the errors in the estimated values of the mass and coupling, we limit our analysis to the case of $SU(2)_F$. In so doing, we shall subtract the contribution of the lowest ground states obtained previously from LSR and work with the $1^{st}$ radial excitation $\oplus$ QCD continuum in a higher range of $t_c-$values.

\begin{figure}[hbt]
\begin{center}
\includegraphics[width=6.0cm]{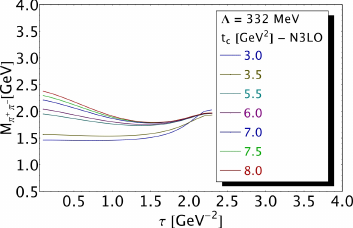}
\includegraphics[width=6.0cm]{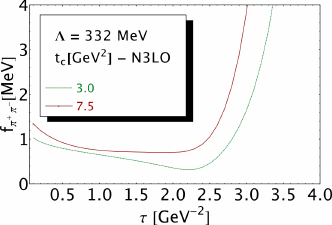}
\caption{\footnotesize The $\tau-$behaviour of the mass and coupling of the 1st radial excitation of the $\pi^+\pi^-$ molecule for different values of $t_c$.} 
\label{fig:pirad}
\end{center}
\end{figure}

\nin As the analysis is performed using the same techniques, we shall illustrate it in the case of the $\pi^+\pi^-$ molecule where, the shape of the mass and coupling in term of the LSR variable $\tau$ for different values of $t_c$ are shown in Fig.\,\ref{fig:pirad}. The optimal values are extracted at the inflexion point $\tau=1.1\,\GeV^{-2}$ for the range of $t_c=3.0\sim 7.5\GeV^{2}$:
\beq
\hspace*{-0.1cm}
M^{(1)}_{\pi^+\pi^-} = 1621(514)\MeV,~f^{(1)}_{\pi^+\pi^-}= 665(338)\keV.
\eeq
The final results are compiled in Table\,\ref{tab:resrad}.
\subsection*{$\lozenge$ Nearby radial excitations effects on the ground states}
\begin{figure}[hbt]
\begin{center}
\includegraphics[width=6.0cm]{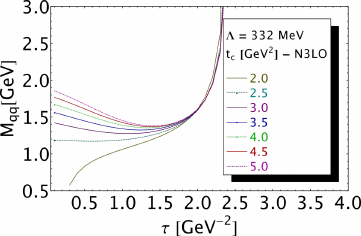}
\includegraphics[width=6.0cm]{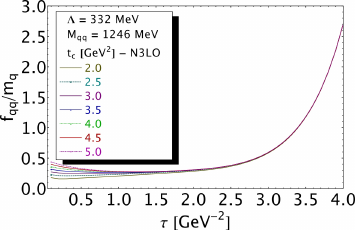}
\caption{\footnotesize The $\tau-$behaviour of the mass and coupling of the $\bar{u}d$ ground state within a "two resonance $\oplus$ QCD continuum" versus the LSR variable $\tau$ and for different values of $t_c$.} 
\label{fig:radeffct}
\end{center}
\end{figure}

\nin We re-estimate the lowest ground state mass and coupling by considering the effect of the $1$st radial excitation. In so doing, we shall use the "two resonances $\oplus$ QCD continuum" for parametrizing the spectral function. As the techniques used for the analysis of the different states are similar, we shall illustrate the study in the case of $\bar{u}d$ ordinary meson shown in Fig.\,\ref{fig:radeffct}. We obtain within a NWA:
\beq
\hspace*{-0.1cm}
M_{\bar{u}d} = 1271(124)\MeV,~f_{\bar{u}d}/\overline{m}_{q}= 243(43)\times 10^{-3}.
\eeq
One can notice that the value agree within the errors with the one in Eq.\,\ref{eq:fresqq}. The effect of the nearby first radial excitation on the determination of the lowest ground state mass from the ansatz "one resonance $\oplus$ QCD continuum" is tiny. 
\section{Comments and Conclusion}
We have presented new results for $\pi^+\pi^-$-like molecule states and improved the existing predictions of the masses and couplings of the ordinary $\bar{q}q$ and $(\overline{qq'})(qq')$ four-quark states. The results with the different sources of the errors are given in Tables\,\ref{tab:resqqmol} to \ref{tab:resrad}.\\
$\circ$ One can notice from the different results that the effects of the SU(3) breaking are only about some few tens of $\MeV$ for the different states.\\
$\circ ~~\sigma/f_0(500)$\\
-- The $R_{P/C}$ condition favours the estimation of the mass of about $1\,\GeV$ but exclude the values around $(0.5-0.6)\MeV$ obtained in the recent literature.\\
-- The constraint on the $\sigma \overline{K}K$ coupling\,\cite{KMSN,MSNW} does not favour the pure $\bar{u}\bar{d}ud$ and $\pi^+\pi^-$ configuration for the $\sigma$.\\
$\circ ~~f_0(980)$\\
-- The $\sigma$ and $f_0(980)$ seems to emerge from a maximal meson-gluonium mixing with $(M_{\bar{q}q},\Gamma_{\pi\pi})=(1229,120)\,\MeV$\,\cite{ANRls} and the light scalar gluonium mass $(M_G,\Gamma_G)=(1070,890)\MeV$\,\cite{SNA}.\\
-- The mass of the $K^+K^-$ molecule and the mean prediction of the four-quark states $M_{\bar{u}\bar{s}ds}=1045(112)\MeV$ are compatible with the $f_0(980)$.\\
$\circ ~~a_0(980)$\\
The $\eta\pi^0$ molecule and the mean of the four-quark states are compatible.\\
$\circ ~~f_0(1370)$\\
This state can be explained by the 1st radial excitation of the $\bar{q}q$ which can mix with the scalar gluonium $M_{\sigma'}=1110(117)\MeV$ to give the large $\pi\pi$ width. It can also be explained by the mix of the radial excitation of the four-quark $M_{\bar{u}\bar{d}ud}=1409(112)\MeV$ with the previous gluonium state.\\
$\circ ~~a_0(1450)$\\
This state is compatible with the 1st radial excitation of the $\bar{u}\bar{s}ud$ four-quark which should be almost degenerated to the $\bar{u}\bar{d}ud$ state.\\
$\circ ~~f_0(1500)$\\
Expected to be a gluonium state from its mass $M_{G'}=1563(141)\MeV$ and from its $U(1)-$like decays $(\eta'\eta, \eta\eta)$\cite{SNA,SNV}. From our results, this state is reproduced by the 1st radial excitation of the four-quark state $M_{\bar{u}\bar{d}ud}=1409(112)\MeV$ which may mix with the previous gluonium state.\\
$\circ ~~f_0(1710)$\\
This state can be reproduced by the 1st radial excitation of the four-quark or/and molecule states.\\
$\circ ~~K^{*}_0(700)$\\
The Breit-Wigner mass of $(840\pm 17)\MeV$ \cite{ANRls} is comparable to the ones of the four-quark and the $K\pi$ molecule.\\
$\circ ~~K^{*}_0(1430)$\\
This state can be reproduced by the ordinary $\bar{u}s$ or/and its 1st radial excitation expected to be around $1400\MeV$.

\begin{table*}[hbt]
\begin{center}
\caption{\scriptsize Sources of errors and values of the masses and couplings of the molecules and ordinary $\bar{q}q$ mesons within NWA. The error due to the truncation of the OPE is estimated by assuming that the next non-calculated term is of the form: $\Delta OPE=(\Lambda^2 \tau)\times (D=6\,\mbox{contribution})$. We take $\vert \Delta \tau \vert \simeq 0.2$. In the case of asymmetric errors, we take the mean value.}
\vspace*{-0.2cm}
\includegraphics[width=16.5cm]{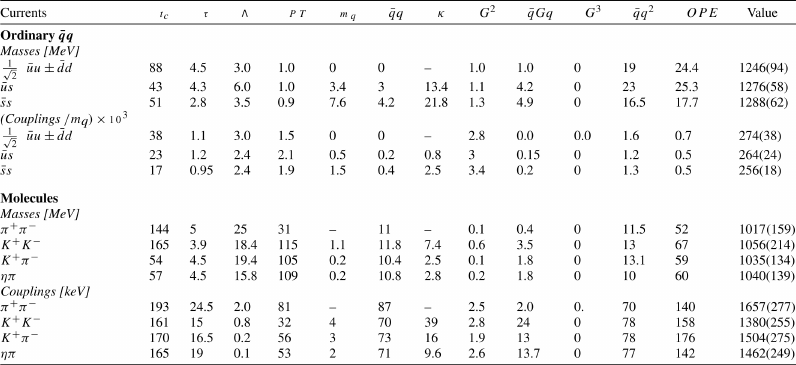}
\label{tab:resqqmol}
\end{center}
\end{table*}
\begin{table*}[hbt]
\begin{center}
\caption{\scriptsize The same caption as for Table\,\ref{tab:resqqmol} but for the four-quark states in the Scalar $\oplus$ Pseudoscalar configurations and for three typical values of $r=1,\,1/\sqrt{2},\,0$.}
\vspace*{-0.2cm}
\includegraphics[width=16.5cm,height=9.8cm]{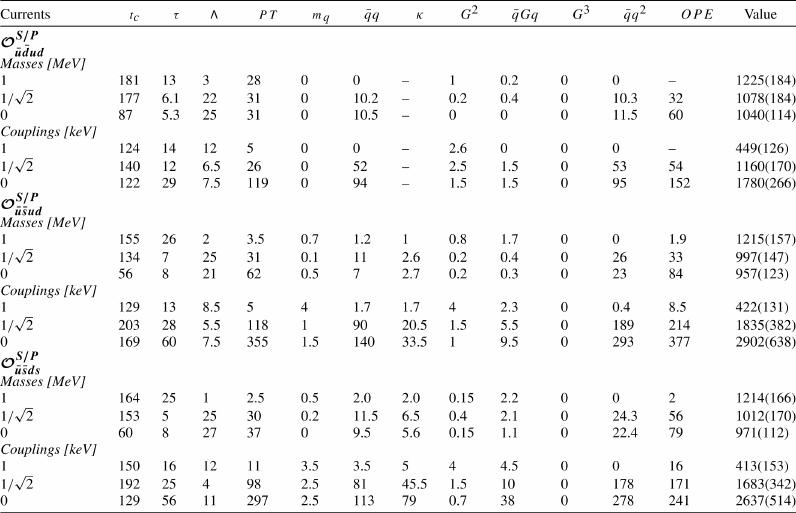}
\label{tab:ressp}
\end{center}
\end{table*}
\begin{table*}[hbt]
\begin{center}
\caption{\scriptsize The same caption as for Table\,\ref{tab:resqqmol} but for the four-quark states in the Vector $\oplus$ Axial-vector configurations and for two typical values of $r=1,\,1/\sqrt{2}$.}
\vspace*{-0.2cm}
\includegraphics[width=16.5cm]{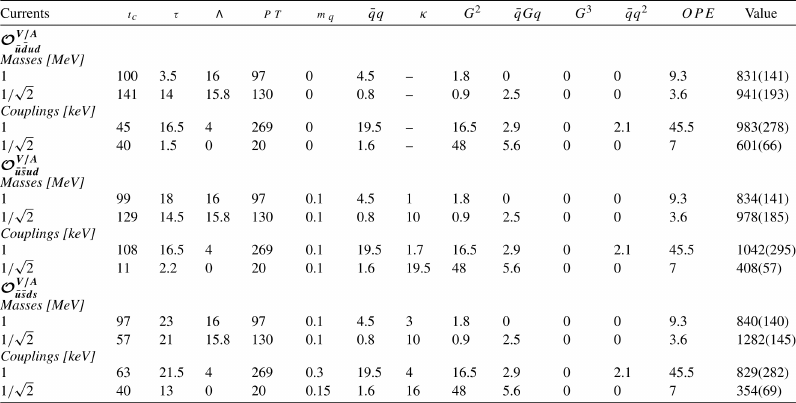}
\label{tab:resva}
\end{center}
\end{table*}
\begin{table*}[hbt]
\begin{center}
\caption{\scriptsize The same caption as for Table\,\ref{tab:resqqmol} but for the $1$st radial excitations of the different assignments. $r=1,\,1/\sqrt{2},\,0$ are typical values of the four-quark mixing of currents.}
\vspace*{-0.2cm}
\includegraphics[width=16.5cm]{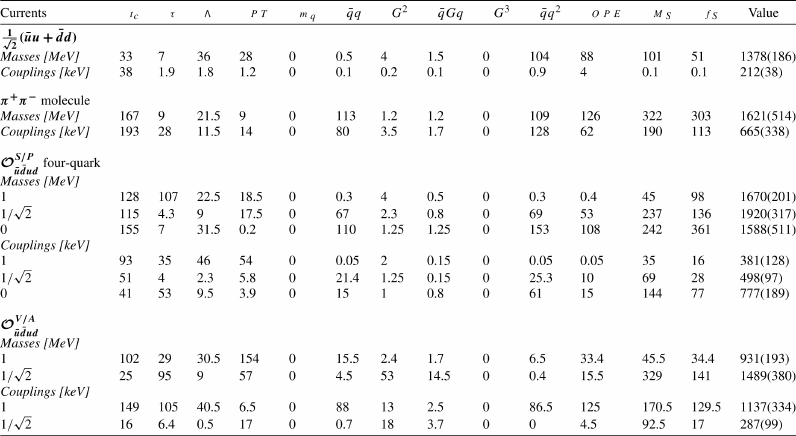}
\label{tab:resrad}
\end{center}
\end{table*}

\end{document}